\documentclass[pdflatex,sn-mathphys]{sn-jnl}
\jyear{2022}

\theoremstyle{thmstyleone}

\theoremstyle{thmstyletwo}

\theoremstyle{thmstylethree}

\begin{document}

\title[Massively degenerate coherent perfect absorber for arbitrary wavefronts]{Massively degenerate coherent perfect absorber for arbitrary wavefronts}

\author[1]{\fnm{Yevgeny} \sur{Slobodkin}}
\equalcont{These authors contributed equally to this work.}

\author[1]{\fnm{Gil} \sur{Weinberg}}
\equalcont{These authors contributed equally to this work.}

\author[2]{\fnm{Helmut} \sur{Hörner}}
\equalcont{These authors contributed equally to this work.}

\author[2]{\fnm{Kevin} \sur{Pichler}}

\author*[2]{\fnm{Stefan} \sur{Rotter}}\email{stefan.rotter@tuwien.ac.at}

\author*[1]{\fnm{Ori} \sur{Katz}}\email{orik@mail.huji.ac.il}

\affil[1]{\orgdiv{Applied Physics Department}, \orgname{Hebrew University of Jerusalem}, \orgaddress{\city{Jerusalem}, \postcode{9190401}, \country{Israel}}}

\affil[2]{\orgdiv{Institute for Theoretical Physics}, \orgname{Vienna University of Technology (TU Wien)}, \orgaddress{\city{Vienna}, \postcode{1040}, \country{Austria}}}

\abstract{One of the key insights in the emerging field of non-Hermitian photonics is that well-established concepts like the laser can be operated in reverse to realize a ‘coherent perfect absorber’ (CPA). While conceptually appealing, such CPAs are limited so far to a single, judiciously shaped wavefront or ‘mode’. Here, we demonstrate how this limitation can be overcome by time-reversing a ‘degenerate cavity laser’, based on a unique cavity that self-images any incident light-field onto itself. Placing a weak, critically-coupled absorber into this cavity, we demonstrate that any incoming wavefront, even a complex and dynamically-varying speckle pattern, is absorbed with close to perfect efficiency in a massively parallel interference process. Moreover, the coherent nature of multi-mode absorption allows us to tune the degree of absorption over a wide range. These characteristics open-up interesting new possibilities for applications in light-harvesting, energy delivery, light control, and imaging.}


\maketitle

\pagestyle{plain}

\section{Introduction}\label{sec1}

The absorption of light is a fundamental process in nature, physics, and engineering, that is central to many important tasks ranging from photo-synthesis to the operation of solar panels and detectors. While light is readily absorbed by thick materials that we perceive as black, thin and weakly absorbing media are inherently far less efficient in capturing incoming radiation and converting it into heat or other forms of energy. A well-known strategy to make even such weakly dissipative substances strongly absorbing is to embed them into a resonant structure \cite{kishino1991resonant,unlu1992theoretical}. At the so-called `critical coupling condition', where the coupling strength to such a resonator is exactly balanced with the internal dissipation, the incoming field gets perfectly absorbed with no energy being back-reflected from the resonator \cite{yariv2002critical}. 
The price one pays for this interferometric enhancement of absorption is a severe restriction on the properties of the incoming field for which this critical coupling condition can actually be satisfied. In the case where only a single incoming channel (mode) is considered, the optical frequency needs to be precisely tuned to the critically coupled resonator resonance frequency \cite{kishino1991resonant}. Generalising the critical coupling condition to multi-channel scattering problems leads to the phenomenon of `coherent perfect absorption' (CPA) for which the incoming wavefront in all available scattering channels needs to be adjusted, in addition to the spectral tuning \cite{chong2010coherent,baranov2017coherent}. In other words, whether it is two laser beams impinging on an absorbing structure \cite{wan2011time,noh2012perfect,Baranov:17} or a complex microwave field hitting a disordered arrangement of obstacles \cite{pichler2019random}, at the critical coupling condition, only a single, suitably adjusted wavefront (or spatial mode) gets `coherently perfectly absorbed'. While this required wavefront adjustment opens up the possibility to control the absorption process interferometrically \cite{baranov2017coherent}, it also comes with the severe limitation that, apart from the correctly matched input wavefront, all of the possibly many other modes are only weakly absorbed due to the different interference patterns they create. In an attempt to overcome this restriction, recent works have shown how to merge two perfectly absorbed modes at a so-called exceptional point, resulting in chiral absorption \cite{sweeney2019perfectly,wang2021coherent,soleymani2022chiral}.

Here, we demonstrate how to entirely eliminate the limitation of the number of perfectly absorbed modes in a CPA. Our design principle for a corresponding multi-mode CPA is based on the insight that `coherent perfect absorption' formally corresponds to the time-reverse of laser emission at the first lasing threshold \cite{noh2012perfect,longhi2010pt}. In the aim to create a device that can perfectly absorb arbitrary combinations of incoming modes interferometrically, one thus needs to time-reverse a laser that emits all of these modes in parallel. Such a laser, indeed, exists, and is known under the name of a `degenerate cavity laser' \cite{arnaud1969degenerate,nixon2013observing,tradonsky2019}. The degeneracy of modes in such a cavity is based on the special feature that the field on either one of the two outer cavity mirrors is self-imaged onto itself after one cavity round-trip. This is realized in a straightforward fashion by placing two lenses in an imaging telescope configuration inside the cavity, ensuring coherent perfect absorption of any combination of modes—regardless of their relative phases. Such a robust absorption mechanism presents an important advantage to a large number of potential applications (see Discussion).

\begin{figure}[t]
\centering{}\includegraphics[width=0.999\textwidth]{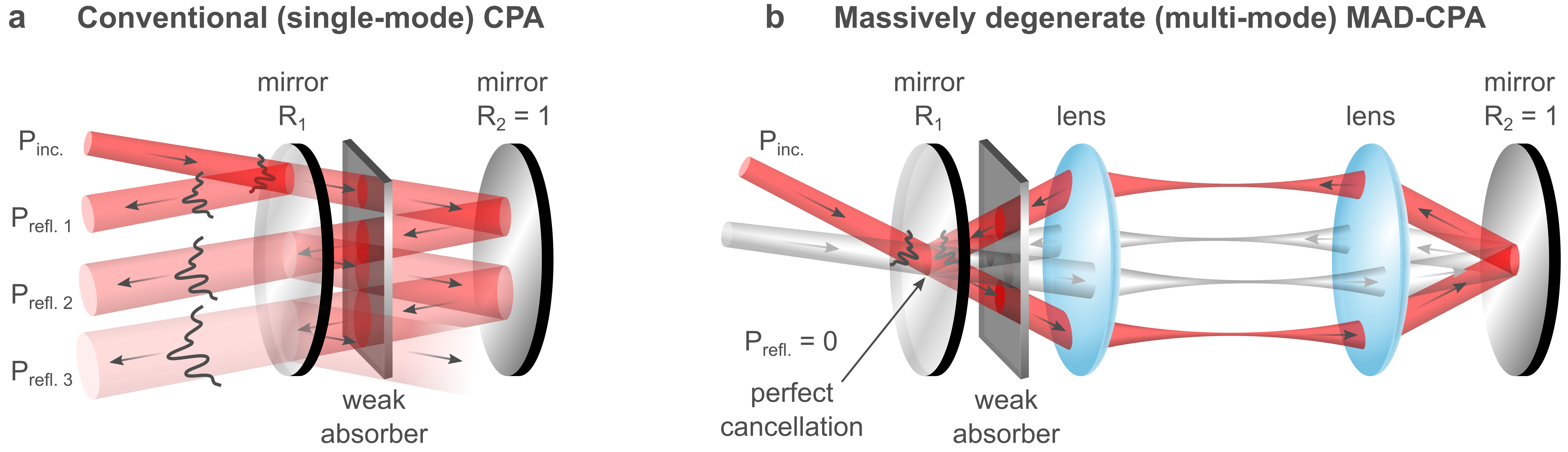}
\caption{\label{fig1:4f_CPA_concept} \textbf{Concept of a massively degenerate coherent perfect absorber (MAD-CPA) for arbitrary wavefronts.} \textbf{a}, A conventional (single-mode) CPA is composed of a weak absorber placed between  two flat mirrors. 
While perfect absorption can be achieved for a normal-incident plane-wave via destructive interference of reflections, any other incoming mode, such as the simple tilted beam shown, results in multiple reflections that cannot destructively interfere.  
\textbf{b}, In contrast, the proposed massively degenerate multi-mode CPA can perfectly absorb any complex incident wavefront. This is achieved by placing the weak absorber in a degenerate (self-imaging) cavity, realized here by a conventional cavity with two lenses in a telescopic arrangement. In such a degenerate cavity, any complex input field impinging on the front cavity-mirror ($R_1$) is self-imaged onto itself after each cavity round-trip: all reflections from the multiple cavity round-trips show perfect destructive interference with the outer reflection from the front cavity-mirror (here shown for two incoming beams at different angles), leading to perfect absorption of light in the weak absorber.}
\end{figure}

The concept of a `massively degenerate CPA' (MAD-CPA) is illustrated in Fig.~\ref{fig1:4f_CPA_concept}, side by side with a conventional single-mode CPA. The simplest conventional CPA is engineered to absorb a plane-wave input at normal incidence, by placing a  critically-coupled absorber between two mirrors. For such an input, all reflections from the multiple cavity round-trips overlap, and destructively interfere when the CPA condition is met. The total reflection is reduced to zero and all the energy is absorbed.
However, for any other input field that is incident at a different angle or in another mode, the reflected fields from the multiple round-trips do not have a spatially identical distribution anymore: their destructive interference is out of sync and perfect absorption cannot be achieved, see Fig.~\ref{fig1:4f_CPA_concept}a. 
In order to realize a CPA that can universally absorb any arbitrary, complex spatial mode, one must assure that all of the resonant cavity reflections coincide and destructively interfere with the non-resonant reflection at the front cavity mirror. This condition is naturally fulfilled in a degenerate (`self imaging') cavity design,  see Fig.~\ref{fig1:4f_CPA_concept}b, which forms the basis for degenerate cavity lasers, studied extensively for their unique lasing properties \cite{nixon2013real,cao2019complex,chriki2021real}. 
Importantly, the self-imaging is maintained for any mode supported by the cavity optics—be it a plane wave at any angle, or a highly complex mode with a complicated wavefront, and even spatially-incoherent fields (see below).

\section{Results}\label{sec2}

\begin{figure} [t]
\centering{}\includegraphics[width=0.999\textwidth]{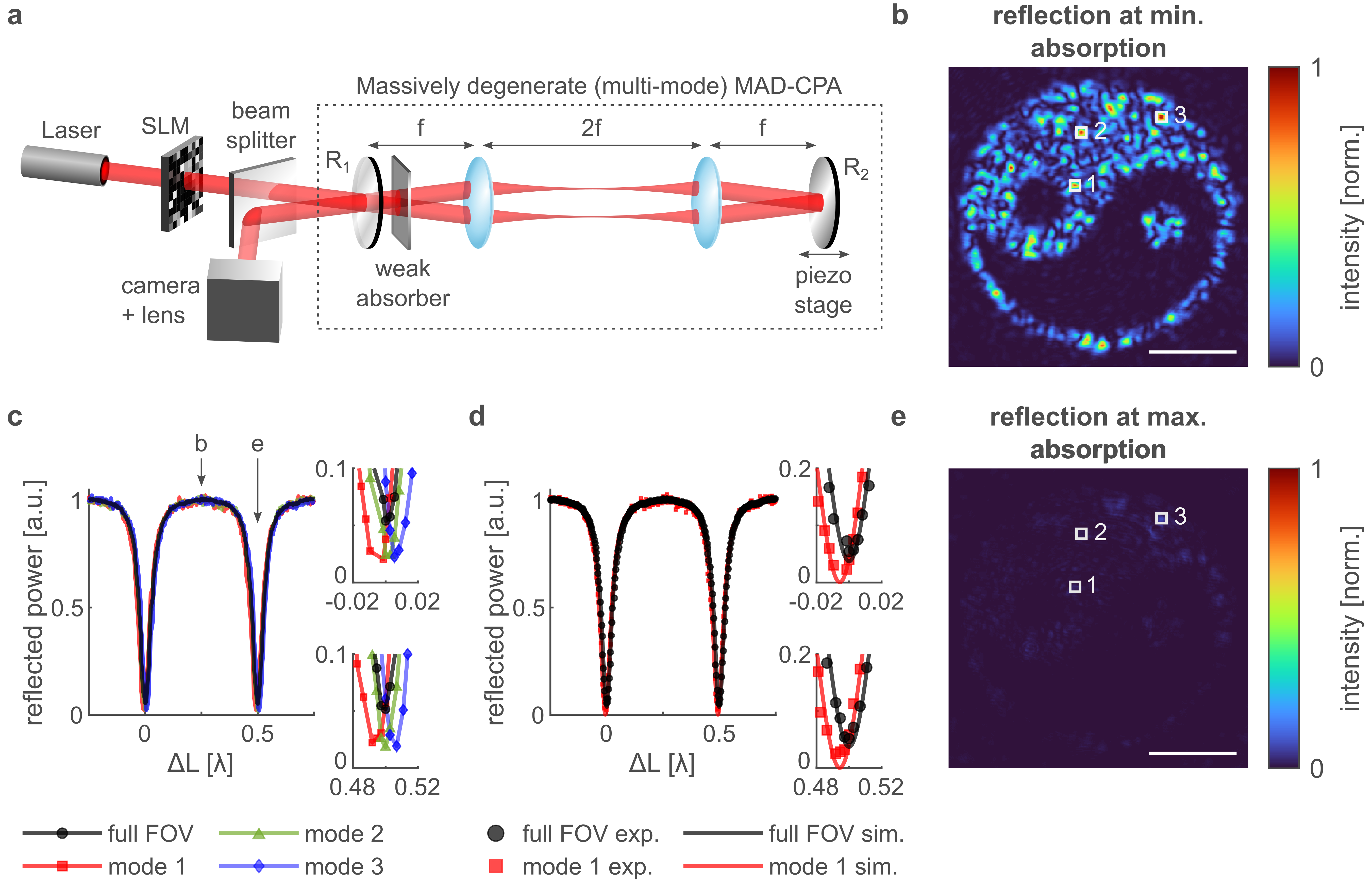}
\caption{\label{fig2:Setup_and_Ying_Yang} \textbf{MAD-CPA  setup and experimental results.} \textbf{a}, Setup: Complex input fields are injected into a degenerate cavity that contains a critically-coupled weak absorber, using a computer-controlled spatial light modulator (SLM), illuminated by a wavelength-stabilized laser. The spatial intensity distribution of the reflected light is measured by a camera. Coherent perfect absorption is achieved by tuning the cavity length to be resonant with the laser wavelength.  
\textbf{b-e}, Results for a complex input field in the form of a speckled Yin-Yang symbol composed of $>1000$ modes. 
\textbf{b}, Measured reflected intensity distribution when the cavity length is tuned for minimal absorption, coherently suppressing absorption (see \textbf{c}), reproducing the incident input field. 
\textbf{c}, Total back-reflected power as a function of the cavity length (black trace). Blue/red/green traces: measured reflected power of three individual localized modes 1/2/3 (speckle grains), marked by white squares in \textbf{b} and \textbf{e}. Insets: zoomed plots of the two minima in \textbf{c}.
\textbf{d}, Same experimental measurements as in \textbf{c} for the total reflected power (black dots) and for the reflected power of one individual mode (red squares), 
together with the numerical prediction for the total reflected power for a 100-modes input (black lines), and a single mode (red line), based on the experimental parameters, with no free parameters (see Methods).  Insets: zoomed plots of the two minima in \textbf{d}.
\textbf{e}, Measured reflected intensity distribution when the cavity length is tuned for maximum absorption, showing near perfect absorption of all input modes. Scale-bars: 1mm.
}
\end{figure}

The experimental setup for the realization and characterization of our MAD-CPA is displayed in Fig.~\ref{fig2:Setup_and_Ying_Yang}a. The central part of this experimental implementation is a degenerate linear cavity composed of a lens-based telescope, that is placed inside a cavity with length of 4-$f$ (with $f$ being the focal length of each lens, see Methods for more details). This resonant cavity features a partially-reflecting mirror at the front (with a reflectivity of $R_1=70\%$) and a near-perfectly reflecting mirror at the back (with $R_2=99.90\%$). 
A weak absorber consisting of a thin color-glass with a single pass transmission of $T_{abs}=85.2\%$ is placed next to the front cavity-mirror. The CPA conditions are met simultaneously for all input modes by adjusting the cavity length to be resonant with the laser wavelength using a piezoelectric translation-stage attached to the back cavity-mirror. Importantly, the coherent nature of absorption in the MAD-CPA allows rapid control including strong suppression of the absorption to values well below the single-pass absorption of the absorber by simple tuning of the cavity length (see below). 
The degenerate CPA is characterized by injecting complex input fields using a spatial light modulator (SLM) illuminated by a collimated beam from a wavelength-stabilized Helium-Neon laser. For each injected complex field we measure the spatial distribution of the reflected light by imaging the front cavity mirror on an sCMOS camera. The very weak light intensity transmitted through the back cavity-mirror is measured by a second camera (not shown).

With the MAD-CPA being the equivalent of a time-reversed degenerate laser at threshold, the three conditions necessary for reaching degenerate coherent perfect absorption are readily deduced: (i) the mirror reflections are equal to the absorption losses (critical coupling): $R_1R_2=T_{abs}^2$; (ii) the cavity is aligned for perfect self-imaging; (iii) the cavity length is resonant with the input wavelength. 
These conditions are met in our experiments by adjusting the reflectivity of the front mirror to the absorber, by carefully aligning the cavity optics (see Methods), and by tuning the cavity length to yield a minimum in the reflected power (see Fig.~\ref{fig2:Setup_and_Ying_Yang}c-e).

To illustrate the versatility of our MAD-CPA setup, we inject into it a highly complex input field, in which more than $1000$ modes coherently form a speckled Yin-Yang symbol. Figure \ref{fig2:Setup_and_Ying_Yang}b-e presents the corresponding experimental results for the light field back-reflected from the MAD-CPA, while tuning the cavity length. As a hallmark of the successful operation of our device, we observe that the reflected power of all the input modes in the Yin-Yang input field is simultaneously minimized when the cavity length is tuned to meet the CPA condition (see Fig.~\ref{fig2:Setup_and_Ying_Yang}c,e). 
The minimum experimental value we reach for the reflectivity is  $\approx 5\%$, indicating that the weak intra-cavity absorber with only $15\%$ absorption in a single pass, now features $>94\%$ absorption for all incoming modes simultaneously. Moreover, each spatially-localized mode (speckle) of the complex field is near perfectly absorbed, reaching a reflected power of $\lesssim 2 \%$  (see Fig.~\ref{fig2:Setup_and_Ying_Yang}c,d). Every localized mode shows similar results to the ones demonstrated previously \cite{wan2011time}, but here for a considerably weaker absorber. When the cavity length is tuned by $\lambda/4$ away from perfect absorption condition, absorption is interferometrically suppressed to values well below the incoherent single-pass transmission of the absorber, providing a modulation-depth of $>50$ for each spatially-localized mode.

The small deviations from the perfect absorption value of $100\%$ are attributed to very small aberrations in the self-imaging cavity optics (on the sub-wavelength scale of $<\lambda/100$), which lead to slightly different cavity round-trip lengths for the different modes (see Fig.~\ref{fig2:Setup_and_Ying_Yang}c, inset, and Supplementary Fig.~\ref{figS1characterisation}). In addition, weak spurious reflections from the cavity lenses' anti-reflection coatings (with $R\approx 0.13\%$ per surface) contribute to this degradation (see Supplementary Fig.~\ref{figS3:lens_reflections}). The influence of such experimental imperfections  are numerically studied in Supplementary section 2, showing very good agreement (without free parameters) down to the level of the absorption curves for each single mode (see Fig.~\ref{fig2:Setup_and_Ying_Yang}d).

\begin{figure}[t]
\centering{}\includegraphics[width=0.999\textwidth]{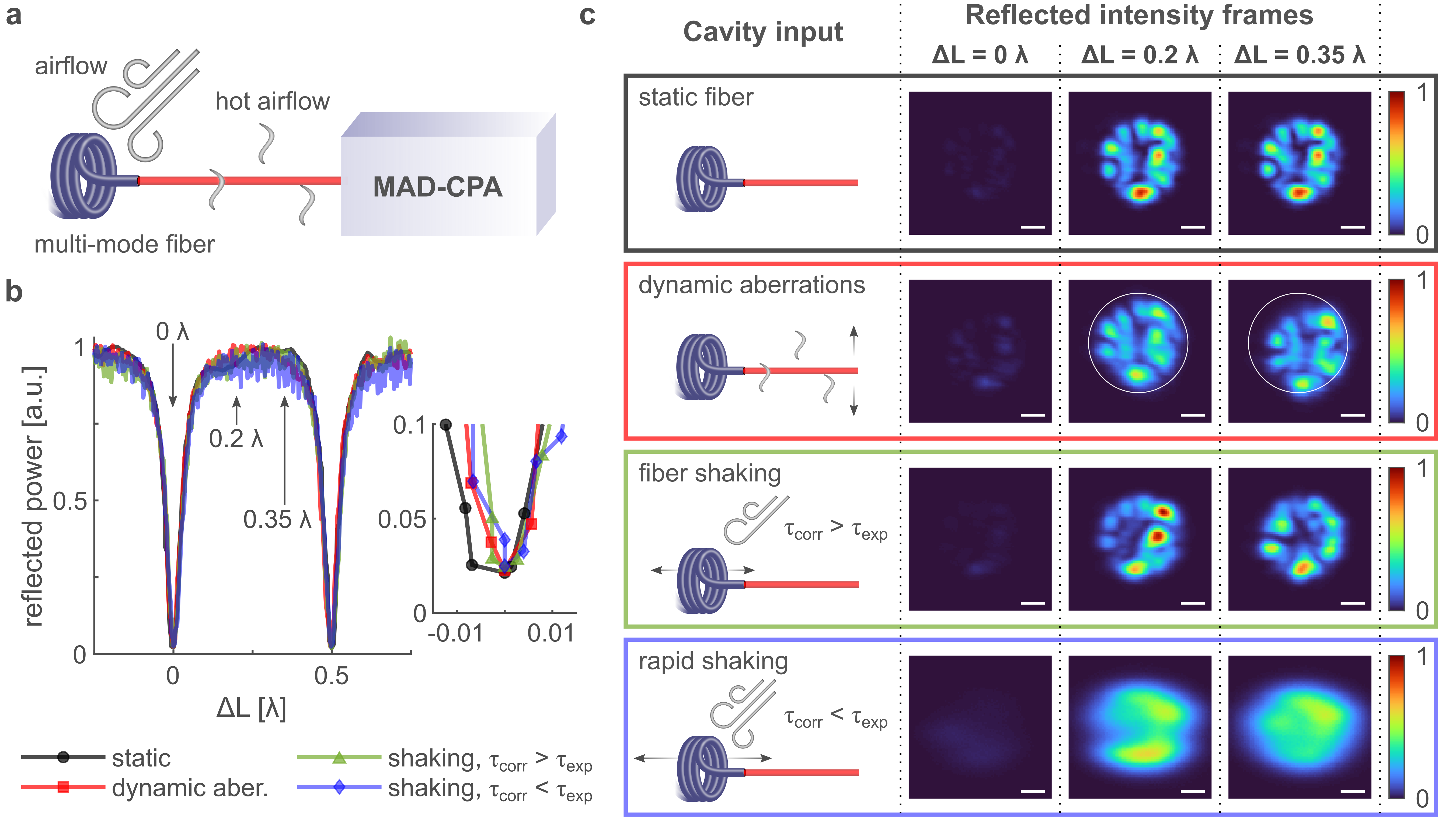}
\caption{\label{fig3:MMF} \textbf{Coherent perfect absorption of rapidly-varying complex fields.} \textbf{a}, Experimental setup: dynamic complex fields are naturally generated by passing the illuminating laser field through a flexible multi-mode fiber (MMF), shaken by strong airflow. The light emanating from the MMF is further subject to atmospheric aberrations generated by passing it through a hot airflow from a heat-gun. \textbf{b}, The total back-reflected power as a function of cavity length in four different experiments. Black trace: static fiber; red trace: static fiber with additional dynamic atmospheric aberrations; green trace: dynamic shaking of the fiber by airflow; blue trace: rapid shaking of the fiber, where the field dynamics are faster than the camera exposure time, effectively demonstrating absorption of spatially-incoherent fields. Inset: zoom of the left minimum in \textbf{b}.
\textbf{c}, Individual camera frames from the four experiments, showing the reflected intensity at maximum absorption ($\Delta L = 0$) and at two differernt cavity lengths  ($\Delta L = 0.2 \lambda$, $\Delta L = 0.35 \lambda$), demonstrating interferometric control of absorption. Scale-bars: 100$\mu m$. 
}
\end{figure}

To further demonstrate the flexibility inherent in the design of the MAD-CPA, we now illustrate its ability to absorb dynamic, rapidly-changing complex random light fields, naturally generated by transmission through flexible multi-mode fibers (MMF) and dynamic atmospheric aberrations (see Fig.~\ref{fig3:MMF}). 
We achieve this by replacing the SLM with a $40\mathrm{cm}$ long MMF (see Methods). The coherent light propagation through the MMF generates complex speckle fields due to the dispersion of the fiber modes (see Fig.~\ref{fig3:MMF}c). To generate not only such a spatial complexity, but also complex temporal dynamics, we rapidly shake the MMF using an external airflow. Moreover, before injecting the speckle fields generated by the MMF into the cavity, we let them propagate through dynamic atmospheric aberrations, generated by a hot air stream from a heat-gun (see Fig.~\ref{fig3:MMF}a). The results of these experiments are shown in Fig.~\ref{fig3:MMF} for different dynamic variations. In all cases, similar near-perfect absorption values are achieved at the CPA conditions—regardless of the input complexity or temporal dynamics (see Fig.~\ref{fig3:MMF}b). Note that the absorption results remain unchanged for temporal variations that are faster than the camera exposure time, representing the absorption of spatially-incoherent light fields—the time-reversed version of a spatially-incoherent lasing field \cite{cao2019complex}. 
Absorption will remain unchanged as long as the bandwidth of the input fields is narrower than the cavity absorption linewidth. This enables perfect absorption of dynamically varying fields as long as the correlation time, $\tau_{corr}$, of the temporal dynamics is longer than the photon-decay time in the cavity, $\tau_c$.
In Fig.~\ref{fig3:MMF}c we also show individual camera frames from these dynamic absorption experiments.

\section{Discussion}\label{sec5}
We present here both the concept and an implementation of a multimode MAD-CPA that enables the coherent perfect absorption of arbitrary incident waves, including dynamically varying complex wavefronts, and spatially incoherent fields. This extension to multiple modes not only removes a central limitation of the CPA-concept, but it is easy to implement and has interesting potential for applications, such as light-harvesting and sensing. In particular, the feature that a weak absorber can perfectly capture any arbitrary incoming field at well-defined frequencies is a unique aspect of our design.

The number of modes that can be supported by the MAD-CPA is limited by the space-bandwidth product of the self-imaging optics \cite{cao2019complex}. Perfect absorption occurs as long as the incoming wavefronts are within the angular acceptance and area of the self-imaging cavity optics \cite{cao2019complex}, and their spectral bandwidth is within the spectral bandwidth of the degenerate cavity. For each localized mode, the absorption bandwidth of the degenerate CPA can be estimated from simple Fabry-Pérot cavity theory \cite{ismail2016fabry}:  $\delta\nu_{\mathrm{FWHM}}= -c\;\mathrm{ln}\left(R_1R_2T_{abs}^2\right)/4\pi L$ , where $c$ is the speed of light in the cavity, and $L$ is the cavity length.

Our experimentally realized degenerate CPA is based on a 4-$f$ linear cavity configuration. However, alternative degenerate cavity designs may offer interesting advantages, such as those making use of a smaller number of optical elements \cite{arnaud1969degenerate}. Self-imaging cavity designs have recently also been proposed for improving the contrast in multi-pass microscopy \cite{juffmann2016multi,klopfer2021continuous}. With these existing concepts being fundamentally different from a degenerate CPA, since no coherence or critical coupling conditions are satisfied, it may be interesting to explore if perfect absorption can also be a useful tool in the context of enhanced sensitivity in microscopy. 

The ability, of a MAD-CPA to enhance or suppress absorption with high contrast for thousands of modes simultaneously, may carry an interesting potential for optical modulation and switching.
Other promising extensions of the MAD-CPA concept include its use as a highly-multimode reflection-less scattering system \cite{sweeney2020theory}, e.g., by replacing the absorber with a scatterer. 
Placing an SLM inside the laser cavity \cite{cao2019complex}, may also allow one to digitally control or compensate absorption and aberrations. 
In view of practical applications, we have studied in our experiments a single-port realization of a degenerate CPA. The presented concept can, however, also be extended to multi-port realizations, e.g., by choosing a partially-reflecting back-mirror ($R_2<1$). In this case, however, input fields need to enter the cavity from both the front and the back-side and additional conditions on the phase and amplitude relations between these fields are required to reach the CPA conditions, similar to a conventional two-port CPA.
	
Last, but not least, we emphasize that our work has focused on the spatial degrees of freedom of the incoming waves. It would be fascinating to study the potential of cavity designs to achieve comparable advances in the spectral domain  \cite{wu2008white,kotlicki2014wideband, kim2016general}, leading potentially to broadband absorption characteristics—a direction in which advances have recently been made using exceptional points and non-linear media \cite{sweeney2019perfectly,wang2021coherent,soleymani2022chiral,Suwunnarat2022}.

\section{Methods}

\textbf{Experimental Setup.} The degenerate optical cavity is realized by a $R=70\%$ partially-reflective mirror (Thorlabs BST10) as the front cavity mirror, and a highly reflecting back cavity mirror (Layertec 104239, $R=99.90\%$). Self-imaging is realized by two $f=75mm$ anti-reflection coated bi-convex lenses (Thorlabs LB1901-A) in a 4-$f$ imaging telescope configuration. The absorber is an AR-coated ionically-colored glass absorber (Thorlabs NE01A-A, $T=85.2\%$). The cavity length is controlled by placing the back cavity mirror on a translation-stage driven by a piezoelectric actuator (Thorlabs NFL5DP20S), having a resolution of $\delta z=0.6nm$, and a step size of $\Delta z = 1.5nm$.

The laser source is a wavelength-stabilized Helium-Neon Laser (Thorlabs HRS015B). The spatial-light modulator is a phase-only SLM (PLUTO-NIR-011). The computer generated holographic fields are generated by Fourier transforming the SLM on the front cavity mirror, by placing the SLM at the back focal plane of a $f=250mm$ bi-convex lens (Thorlabs LB1056-B). The front cavity-mirror is placed at the lens' front focal plane.
In the experiments of Fig.~\ref{fig3:MMF}, the MMF is a $40\mathrm{cm}$-long, $0.1$ NA fiber, with a core diameter of $25\mu m$ (Thorlabs FG025LJA). The fiber is 4-$f$ imaged on the front cavity-mirror by a telescope with a magnification of $\times13.7$ (not shown in Fig.~\ref{fig1:4f_CPA_concept}).

To measure the back-reflected light, a 50:50 pellicle beamsplitter (Thorlabs BP150) was placed in front of the cavity front mirror. An sCMOS Camera (Andor Zyla 4.2) with 16 bits digitization and a pixel well-depth of 30,000 electrons measured the image of the reflected field at the cavity front mirror surface, which was imaged with a magnification of $\approx 1$ on the camera by an $f=125mm$ lens  (Thorlabs LB1904-B).

Before the experimental measurements, the cavity was aligned to meet the CPA conditions by adjusting the cavity elements position and angle, such that the total reflected power as measured by the camera is minimized, i.e., that the optimal destructive-interference occurs over the entire imaged field-of-view. 

To validate that no significant optical power is transmitted through the high-reflection back cavity-mirror ($R_2=99.90\%$), we measured the transmission through the cavity back-mirror with the same sCMOS camera while the cavity length was scanned. The power transmitted through the cavity back-mirror was not higher than $0.8\%$ of the incident power (see Supplementary Fig.~\ref{figS4:power_callibration}).\\

\textbf{Numerical simulation.} In order to establish a theoretical foundation for the degenerate CPA concept, and to estimate the influence of deviations from an optimal configuration on the performance of the CPA, a numerical simulation model was developed. 
The model and simulations were implemented using scalar Fourier optics theory to calculate the reflection matrices of various degenerate CPA designs. The details of the theoretical model and of the numerical simulations are described in detail in Supplementary Information Section S2.
Specifically, the numerical simulations focused on various deviations from perfect conditions, and were used as a design tool for our experiments. The studied effects included deviations such as mirror tilts and displacement, positions of lenses, and reflections from lenses and mirrors. The numerical simulations including spurious intra-cavity reflection from the surfaces of lenses were used to provide the fit curve in Fig.~\ref{fig2:Setup_and_Ying_Yang}d.

\section{Acknowledgments}
We thank Sylvain Gigan, Yaron Bromberg, and Matthias Kühmayer for helpful discussions. 
This project was supported by: H2020 European Research Council (101002406), Israel Science Foundation (1361/18), National Science Foundation (1813848); Austrian Science Fund (FWF, P32300). The computational results presented were achieved using the Vienna Scientific Cluster (VSC).

\section{Author contributions}
O.K. proposed the project and conceptualized it with S.R.; Y.S., G.W., and O.K. designed the experimental setup. Measurements and data analysis were carried out by Y.S. and G.W., under the supervision of O.K. Theoretical calculations and numerical simulations were carried out by H.H. with input from K.P., Y.S., G.W., and O.K., under the supervision of S.R. All authors contributed to the writing of the manuscript.

\backmatter

\bibliography{main}

\newpage

{\centering\part*{Supplementary material}}

\setcounter{section}{0}
\section{Characterization of multi-mode absorption as a function of cavity length}

\begin{figure}[b]
\renewcommand{\thefigure}{S1}
\centering{}\includegraphics[width=0.99\textwidth]{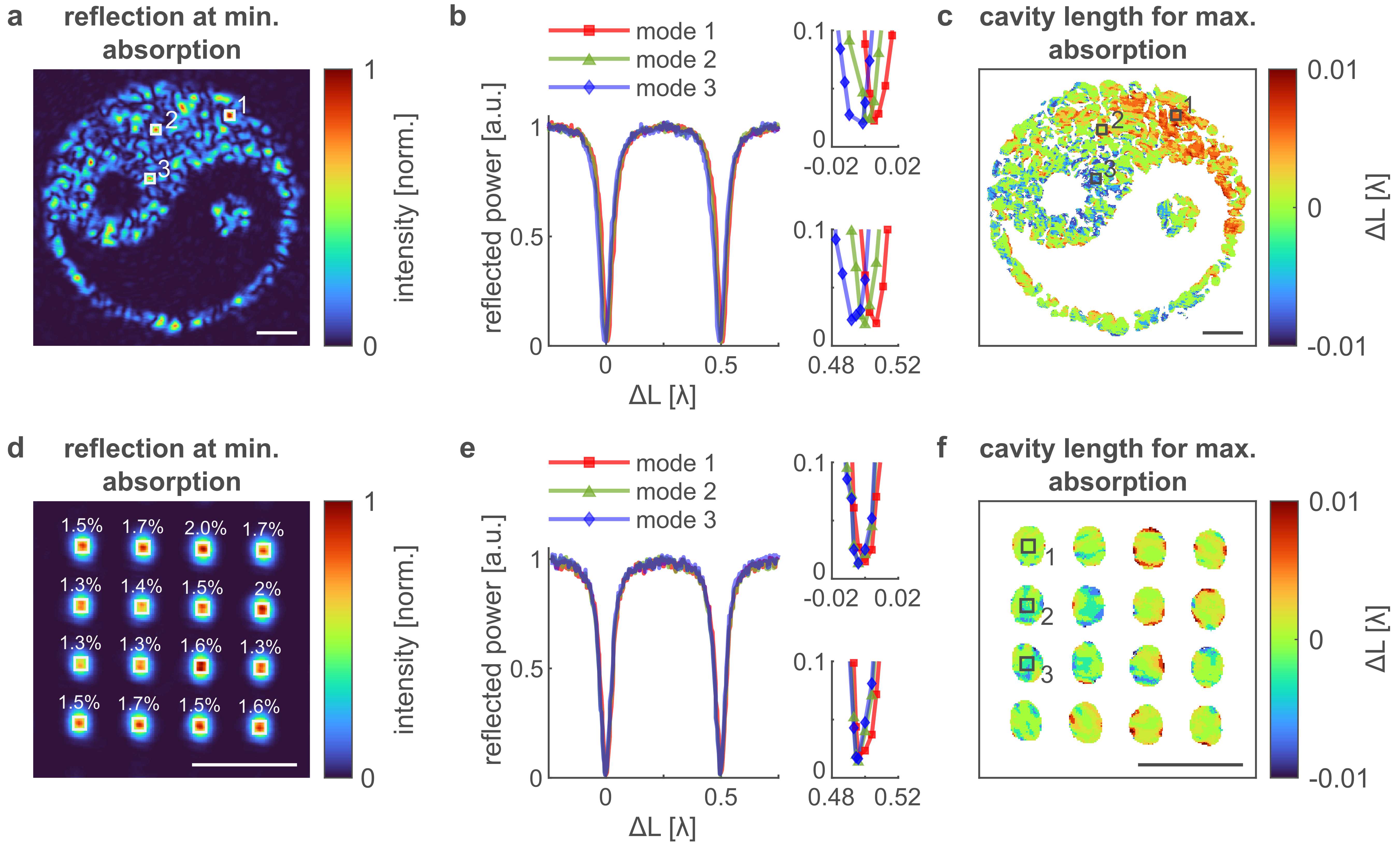}
\caption{\label{figS1characterisation} \textbf{Spatial characterisation of multi-mode absorption as a function of cavity length.} 
\textbf{a}, Measured reflected intensity when the cavity length is tuned for minimal absorption, reproducing the injected input field (same as Fig.~\ref{fig2:Setup_and_Ying_Yang}b). 
\textbf{b}, Measured back-reflected power of three individual localized modes 1/2/3 (speckle grains), marked by white squares in \textbf{a}. Insets: zooms of the two minima in \textbf{b}. Each mode reaches maximum absorption at a slightly different cavity length.
\textbf{c}, Cavity length for maximum absorption, displayed for each pixel. All cavity length shifts are calculated relative to the cavity length where the full field of view absorption is maximal. 
\textbf{d}, Same as \textbf{a}, when injecting an input field pattern consisting of a lattice of 16 diffraction-limited spots. The stated numerical values represent the ratio of the reflected power minima and maxima of each spot.
\textbf{e-f}, same as \textbf{b-c}, for the input displayed in \textbf{d}. Scale-bars: 0.5mm.}
\end{figure}

To provide a more detailed characterization of the performance of our experimental implementation of a MAD-CPA, Fig.~\ref{figS1characterisation} displays a more detailed analysis of the absorption at each transverse spatial position in the cavity input, for both the Yin-Yang input of Fig.~\ref{fig2:Setup_and_Ying_Yang} (Fig.~\ref{figS1characterisation}a-c), and an additional experiment conducted with the same setup and an input field composed of 16 focused diffraction-limited spots spread over a smaller field of view than Fig.~\ref{figS1characterisation}a (Fig.~\ref{figS1characterisation}d-f).

As seen also in Fig.~\ref{fig2:Setup_and_Ying_Yang}c, due to small imperfections in the self-imaging optics, while almost all modes reach very low values for reflection (i.e., very high coherent absorption values), not all of the modes reach maximum absorption at exactly the same cavity length (Fig.~\ref{figS1characterisation}b). To study the origin of these slight mismatches Fig.~\ref{figS1characterisation}c,f shows maps of the cavity length for which each pixel in the reflected field reaches maximum absorption. In the large field of view of Fig.~\ref{figS1characterisation}c the cavity length for maximum absorption can differ by up to $0.01\lambda$, whereas they are appreciably smaller for the small field of view of Fig.~\ref{figS1characterisation}f. 
The measured spatial distributions of the path length differences display a spatially continuous profile that is characteristic of small aberrations in the imaging optical setup. We attribute these to the spherical lenses used for imaging and the limitations in the alignment accuracy.
To complement this study, we have displayed in Fig.~\ref{figS1characterisation}d the minimum reflection values for each of the 16 diffraction-limited spots of the input fields,
where the intensity of each spot is integrated over a window consisting of 81 pixels around each spot center (integration window presented as a white rectangle in Fig.~\ref{figS1characterisation}d and a black rectangle in Fig.~\ref{figS1characterisation}f).

\section{Numerical study of sensitivity to system parameters}
\subsection{Basic simulation method}

In order to design our experiments, we performed a number of computer simulations using common scalar Fourier Optics methods (see, e.g., \cite{Stark1982}), which allowed an in-depth parameter study of the MAD-CPA design. In this section, we will provide details on these simulations and present their main results.

Assuming that the cavity stretches along the $z$-axis, scalar Fourier Optics allows expressing a monochromatic light field at a given axial ($z$) position as a complex-valued function $U(x, y; z)$ that contains the amplitudes and the phases of a coherent electromagnetic field at arbitrary positions on the $xy$-plane.  

Let $A(k_x, k_y; z)$ be the 2D spatial Fourier transform (i.e., the angular spectrum) of $U(x, y; z)$. Then, light propagation over a distance $\Delta z = z_2 - z_1$ can be expressed with the help of the Fresnel transfer function (\ref{eq:fresnel}):

\begin{equation}
    A\left( k_x, k_y; z_2 \right) = A\left(k_x, k_y; z_1 \right) e^{i k \Delta z} e^{-i \frac{\lambda \Delta z}{4 \pi} \left( k_x^2 + k_y^2 \right)}. \label{eq:fresnel}
\end{equation}\\
Expressions like this, as well as those describing the effect of optical components, can be simulated very efficiently using the Fast Fourier Transform (FFT) \cite{Schmidt2010,Voelz2011}. The function $U(x, y; z)$ is then replaced by a matrix $\mathbf{U}_i$, where each complex-valued entry represents amplitude and phase at a certain $xy$-position, and the index $i$ represents the $i$\nobreakdash-th axial $z$-position for which the light-field is to be calculated. Accordingly, the angular spectrum $A(k_x, k_y; z)$ is described by a matrix $\mathbf{A}_i$, containing the complex-valued amplitudes of all FFT base-functions of the light-field at the $i$-th axial $z$-position.

As a first step, we performed a parameter study for the effects of the various experimental parameters, such as tilts and shifts of optical elements, and deviations from perfect critical coupling. To this end, we performed a large number of computationally-efficient simulations where we neglected the small spurious reflectivity of the lenses’ anti-reflection coated surfaces. In these simulations, we simulated the single round-trip propagation through the cavity by dividing it into $i=1 \ldots N$ steps, where in each step, a single optical element or propagation between elements is taken into account. To simulate the result of the infinite number of cavity round-trips, we expressed  the effect of the $i$\nobreakdash-th optical element in the cavity, or the propagation between elements, by a  transmission-matrix $\mathbf{T}_i$.

After converting the angular spectrum matrix $\mathbf{A}_i$ (which expresses the light field at the $i$-th axial $z$-position) into a vector $\mathbf{a}_i$ (containing all the entries of $\mathbf{A}_i$ in the required order for a matrix-vector-multiplication with $\mathbf{T}_i$), one can calculate the field after the $i$-th propagation step as $\mathbf{a}_{i+1}=\mathbf{T}_i \mathbf{a}_i$. This allows the calculation of the transmission-matrix for a \textit{single} roundtrip through the cavity, $\mathbf{T}_{\mathrm{srt}}$, by multiplying all corresponding transmission-matrices: $\mathbf{T}_{\mathrm{srt}}=\prod_i{\mathbf{T}_i}$.

Note that this method is per se \textit{not} able to calculate the effect of the infinite number of round-trips in the two-mirrored cavity. However, with $r$ and $t$ being the (complex-valued) reflection- and transmission-coefficients of the partially-reflecting front cavity mirror, respectively, the cavity reflection-matrix $\mathbf{R}_{\mathrm{cav}}$ can be expressed as $\mathbf{R}_{\mathrm{cav}}=r\mathbf{I}+t^2 \mathbf{T}_{\mathrm{srt}}\left( \mathbf{I} + r\mathbf{T}_{\mathrm{srt}} + r^2 \mathbf{T}_{\mathrm{srt}}^2 + \cdots\right)$, where $\mathbf{I}$ represents the unity matrix and the cavity back mirror is assumed to be perfect. By generalizing the scalar infinite geometric series formula to matrix calculation, this can be expressed as $\mathbf{R}_{\mathrm{cav}}=r \mathbf{I} + t^2 \mathbf{T}_{\mathrm{srt}}\left( \mathbf{I} - r \mathbf{T}_{\mathrm{srt}} \right)^{-1}$.\\

\subsection{Results}
\subsubsection{Deviations from optimal system parameters}

\begin{figure}[t]
\renewcommand{\thefigure}{S2}
\centering{}\includegraphics[width=0.99\textwidth]{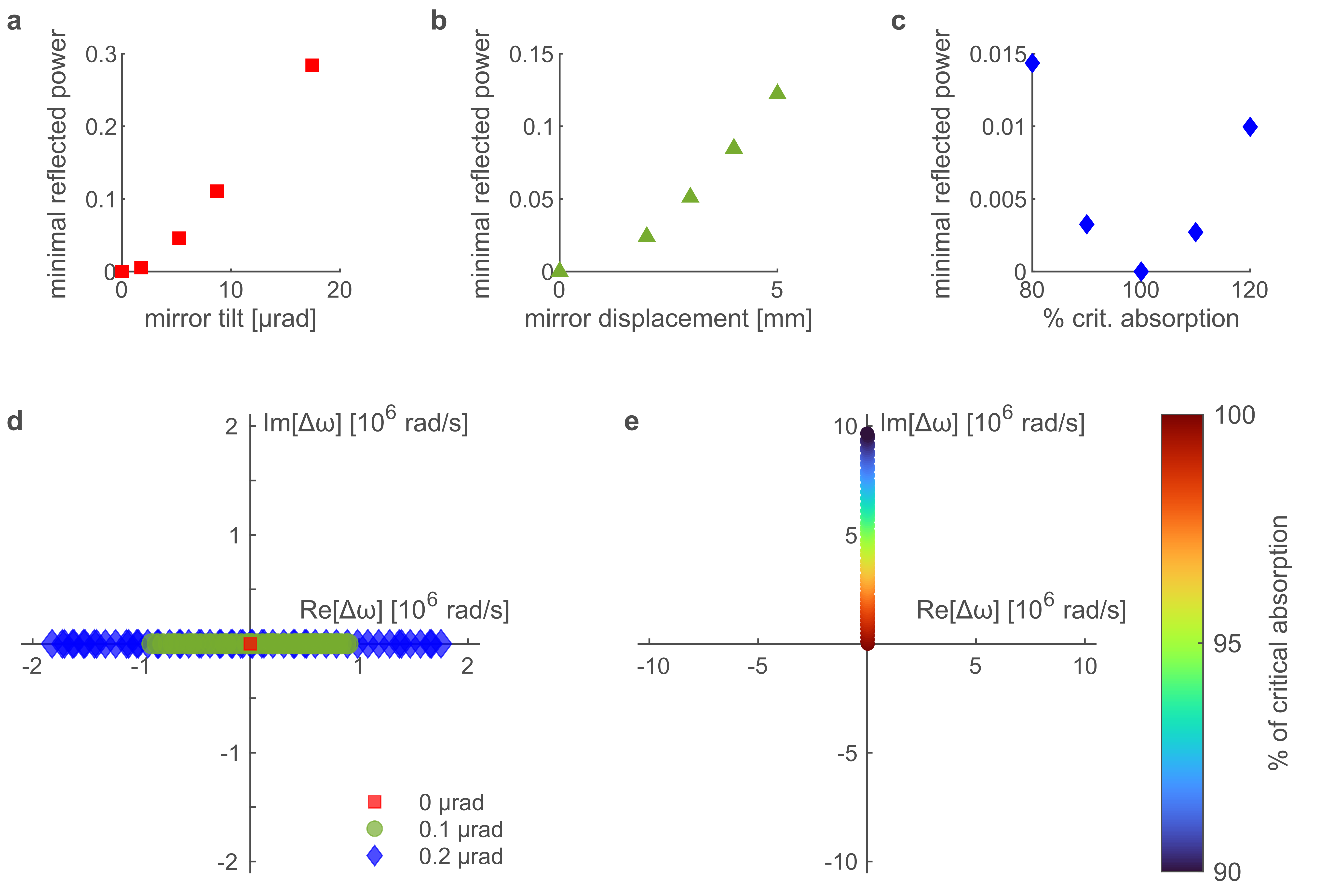}
\caption{\label{figS2:simulaiton} \textbf{Numerical study of CPA performance under deviations from optimal configuration.} 
\textbf{a}, Effect of tilting the CPA rear mirror on the minimum reflected (i.e. maximum absorbed) power.
\textbf{b}, Effect of displacing the CPA rear mirror  on minimum reflected power.
\textbf{c}, Effect of deviations from the critical absorption on the minimum reflected power.
\textbf{d}, The zero-points of the CPA reflection matrix spread out on the real axis in the complex frequency plane when the rear cavity mirror is slightly tilted.
\textbf{e}, The zero-points of the CPA reflection matrix move updwards in the complex frequency plane when the absorption becomes sub-critical.
Simulated input field is a circular random speckle field with $3.1 \mathrm{mm}$ diameter, containing $\sim 1000$ modes. Simulated field of view: $3.56\mathrm{mm} \times 3.56 \mathrm{mm}$, $f = 75 \mathrm{mm}$, $\lambda = 633\mathrm{nm}$, $R_1=70\%, R_2=100\%$, lenses are simulated as thin spherical lenses.
}
\end{figure}

Using the method described above, we calculated the effect of deviations in the positions of the rear cavity mirror,  lenses, deviations from the critical absorption, and the effect of a small tilt in one of the cavity mirrors. From all the before-mentioned deviations, the 4f-cavity-CPA is most sensitive to mirror tilts. Tilting the rear mirror by more than $5$\,\textmu rad already leads to a noticeable deterioration of the CPA (Fig.~\ref{figS2:simulaiton}a). 

Other deviations in parameters have been found to also have noticeable effects, but are easily experimentally maintained. For example, even if the rear mirror is moved out of position by $2 \mathrm{mm}$ in $z$-direction, the minimum reflectivity still only increases to $2.4 \%$ (Fig.~\ref{figS2:simulaiton}b). We leveraged this insensitivity for rear-mirror displacements for the practical implementation of our experiment, as it enabled us to simply shift the rear mirror using a piezoelectric translation stage to reach perfect absorption or quickly tune the cavity length for absorption suppression.

Further simulations have been performed to determine the 4f-CPA's sensitivity against deviations from optimal absorption. As is displayed in Fig.~\ref{figS2:simulaiton}c, the 4f-CPA with the chosen experimental parameters is quite insensitive to both under-critical as well as to over-critical values of absorption.

\subsubsection{Zero-points of the reflection-matrix}
In addition to the above-mentioned simulations, we also numerically explored the behaviour of the zeros of the CPA reflection matrix $\mathbf{R}_{\mathrm{cav}}$. If the 4f-cavity had neither a gain-element nor a loss-element, we would expect $\mathbf{R}_{\mathrm{cav}}$ to be unitary, and hence to have zeros in the upper half, and poles in the lower half of the complex plane of angular frequency $\omega$. However, with just the critical attenuation, we expect CPA behaviour, and all zeros to be situated exactly at the real axis \cite{li_random_2017,fyodorov_distribution_2017}. In our simulation, we took the critical attenuation as a starting point, and then reduced the attenuation step by step towards under-critical attenuation. As expected, the more the attenuation was reduced, the more the degenerate zero-eigenvalues moved upwards in the complex frequency plane in our simulations (see Fig.~\ref{figS2:simulaiton}e).

In a similar simulation, where the total reflective mirror is slightly tilted out of its position perpendicular to the $z$-axis, another effect becomes visible: the single, degenerate zero-point splits up into multiple zero-points, all of which spread on the real axis (see Fig.~\ref{figS2:simulaiton}d).

\subsection{Refined simulation method including spurious reflections from lens facets}

In order to match the experimental results of $\sim94\%$ absorption over the entire field-of-view, the simulations were required to include the effect of residual reflections of the lenses' surfaces used in our experiments. While the lenses used in our experiments have an anti-reflection coating designed for the laser wavelength, there is a residual $0.13\%$ reflection for each of the lenses facets, which, due to the multiple cavity roundtrips, results in a non-negligible effect on CPA performance.
To this end, a more refined (and computationally intensive) simulation approach was chosen. In these simulations, instead of using (one-way) transmission-matrices, the respective $i$-th propagation distance, or $i$-th optical element, was represented by a scattering matrix $\mathbf{S}_i$, such that:

\begin{equation}
    \left( \begin{array}{ll} \mathbf{a}^{\mathrm{left}}_{\mathrm{out}} \\ 
    \mathbf{a}^{\mathrm{right}}_{\mathrm{out}}  \end{array} \right)_{i+1} = 
    \mathbf{S}_i 
    \left( \begin{array}{ll} \mathbf{a}^{\mathrm{left}}_{\mathrm{in}} \\ 
    \mathbf{a}^{\mathrm{right}}_{\mathrm{in}}  \end{array} \right)_i = 
    \left( \begin{array}{cc} \mathbf{R}_i & \mathbf{T}_i \\ \mathbf{T}_i & \mathbf{R}_i \end{array} \right)
    \left( \begin{array}{ll} \mathbf{a}^{\mathrm{left}}_{\mathrm{in}} \\ 
    \mathbf{a}^{\mathrm{right}}_{\mathrm{in}}  \end{array} \right)_i \label{eq:smatrix}
\end{equation}\\
with $\mathbf{R}_i$ and $\mathbf{T}_i$ being the according reflection- and transmission-matrices, calculated by using the Fourier Optics methods explained before. Note that we do not distinguish between left-to-right and right-to-left reflection- and transmission-matrices in the definition of the scattering matrix in eq. (\ref{eq:smatrix}). This is justified as long as all simulated elements are left-right-symmetric, which is indeed the case not only for the propagation in free space and within the absorber, but also for both mirrors and lenses, which are bi-convex and have the same curvature on either side. For simulating the spurious reflectivity of both the lenses' convex outside facets, as well as the concave inside facets, we calculated the $\mathbf{R}_i$ reflection-matrices as a superposition of the outside and the inside reflection effects.  

In the next step, all of the calculated scattering matrices $\mathbf{R}_i$ are converted into multiport transfer-matrices $\mathbf{M}_i$ as explained in \cite{Frei2008}, such that:

\begin{equation}
    \left( \begin{array}{ll} \mathbf{a}^{\mathrm{left}}_{\mathrm{out}} \\ 
    \mathbf{a}^{\mathrm{left}}_{\mathrm{in}}  \end{array} \right)_{i+1} = 
    \mathbf{M}_i 
    \left( \begin{array}{ll} \mathbf{a}^{\mathrm{right}}_{\mathrm{in}} \\ 
    \mathbf{a}^{\mathrm{right}}_{\mathrm{out}}  \end{array} \right)_{i}
\end{equation}\\

The effect of the total 4f-cavity (including the effect of an infinite number of round-trips, and now also including the effects of residual lens reflections) can then be expressed by simply multiplying all transfer-matrices in the according order $\mathbf{M}_{\mathrm{cav}}=\prod_i{\mathbf{M}_i}$. After back-converting the total resulting transfer-matrix $\mathbf{M}_{\mathrm{cav}}$ into the corresponding scattering matrix $\mathbf{S}_{\mathrm{cav}}$ \cite{Frei2008}, one can get the cavity’s total reflection matrix $\mathbf{R}_{\mathrm{cav}}$ by extracting the top-left-quadrant sub-matrix of $\mathbf{S}_{\mathrm{cav}}$. 

This method has, however, two challenges: firstly, the scattering- and transfer matrices are four times the size of the corresponding transmission matrices in the previous approach, which results in the requirement of accordingly larger computing resources.

Even more challenging is the fact that the reflections on the convex lenses’ surfaces result in ever-ongoing radially diverging wavefronts in the $xy$-plane. This is a challenge because of the cyclic nature of FFT: wavefronts leaving the inside of the cavity in the radial direction in physical reality do not automatically do so in the FFT Fourier Optics simulation. To solve this problem, we had to embed the $xy$-observation plane into a much larger simulation grid, on which we simulate a round, hyper-gaussian aperture to dampen the outgoing wavefronts away. Due to these challenges, we could not fully simulate the experimentally realised $\sim 1000$ modes, $3.1 \mathrm{mm}$ diameter speckle field. Instead, we simulated the effect of $100 - 500$ modes random speckle fields with diameters in the magnitude of $1 \mathrm{mm}$, similar to the input fields presented in Fig.~\ref{fig3:MMF} of the main text. 

\subsubsection{Results}
\begin{figure}
\renewcommand{\thefigure}{S3}
\centering{}\includegraphics[width=0.99\textwidth]{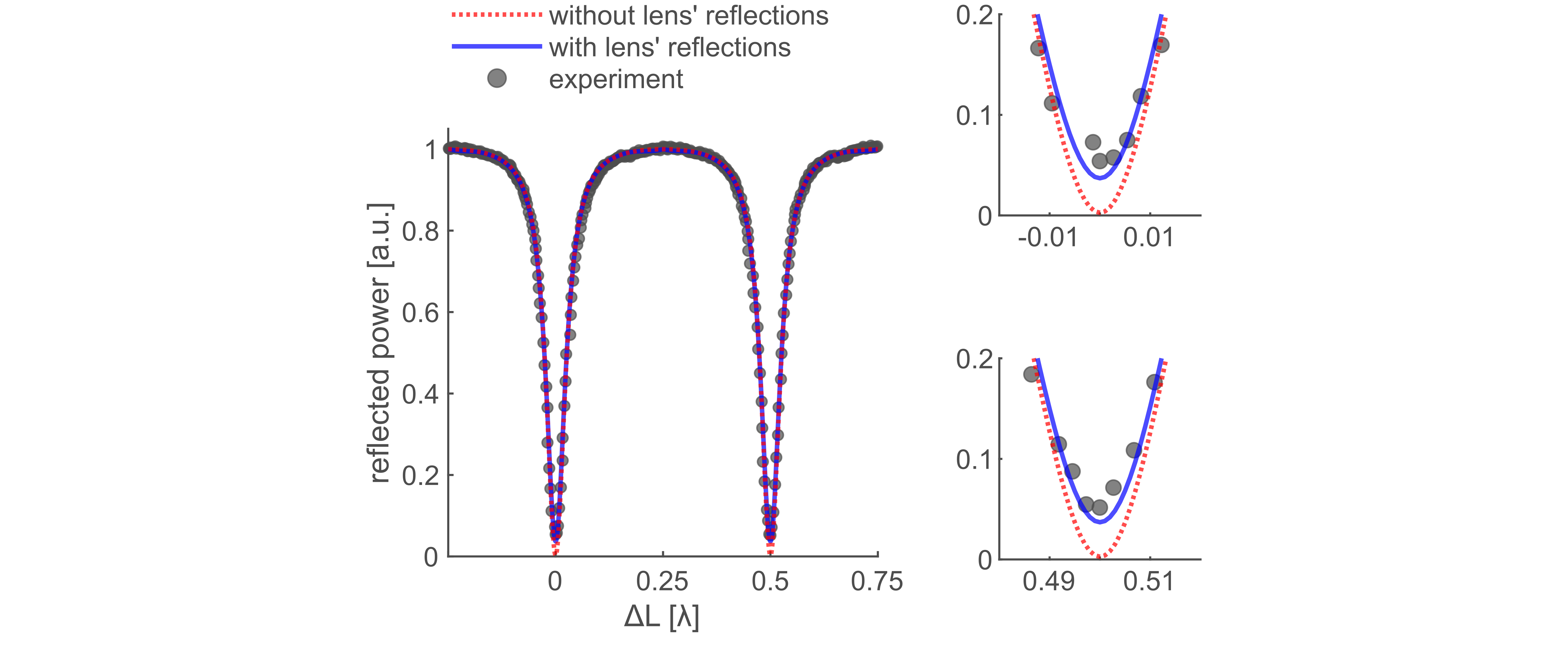}
\caption{\label{figS3:lens_reflections}\textbf{The effects of reflections from lenses' facets on MAD-CPA performance.} Total back-reflected power as a function of the cavity length as measured experimentally (black circles), and for two simulation runs, with and without residual ($0.13 \% $) reflections from the cavity lenses facets (solid blue and dotted red, respectively). The numerical results have no free parameters. Insets: zoomed plots of the two minima.}
\end{figure}

Using the method described above, we were able to reproduce a good match to the experimental results by simply using the stated values for the lenses surfaces reflection ($0.13 \%$), and the effective $14.2 \%$ power absorption of the absorber (Fig.~\ref{figS3:lens_reflections}). These simulations point to the fact that the small residual reflection from the lens' facets is one of the dominant factors in the deviation from $100\%$ perfect coherent absorption.

\section{Experimental measurement of transmitted power}

\begin{figure}
\renewcommand{\thefigure}{S4}
\centering{}\includegraphics[width=0.49\textwidth]{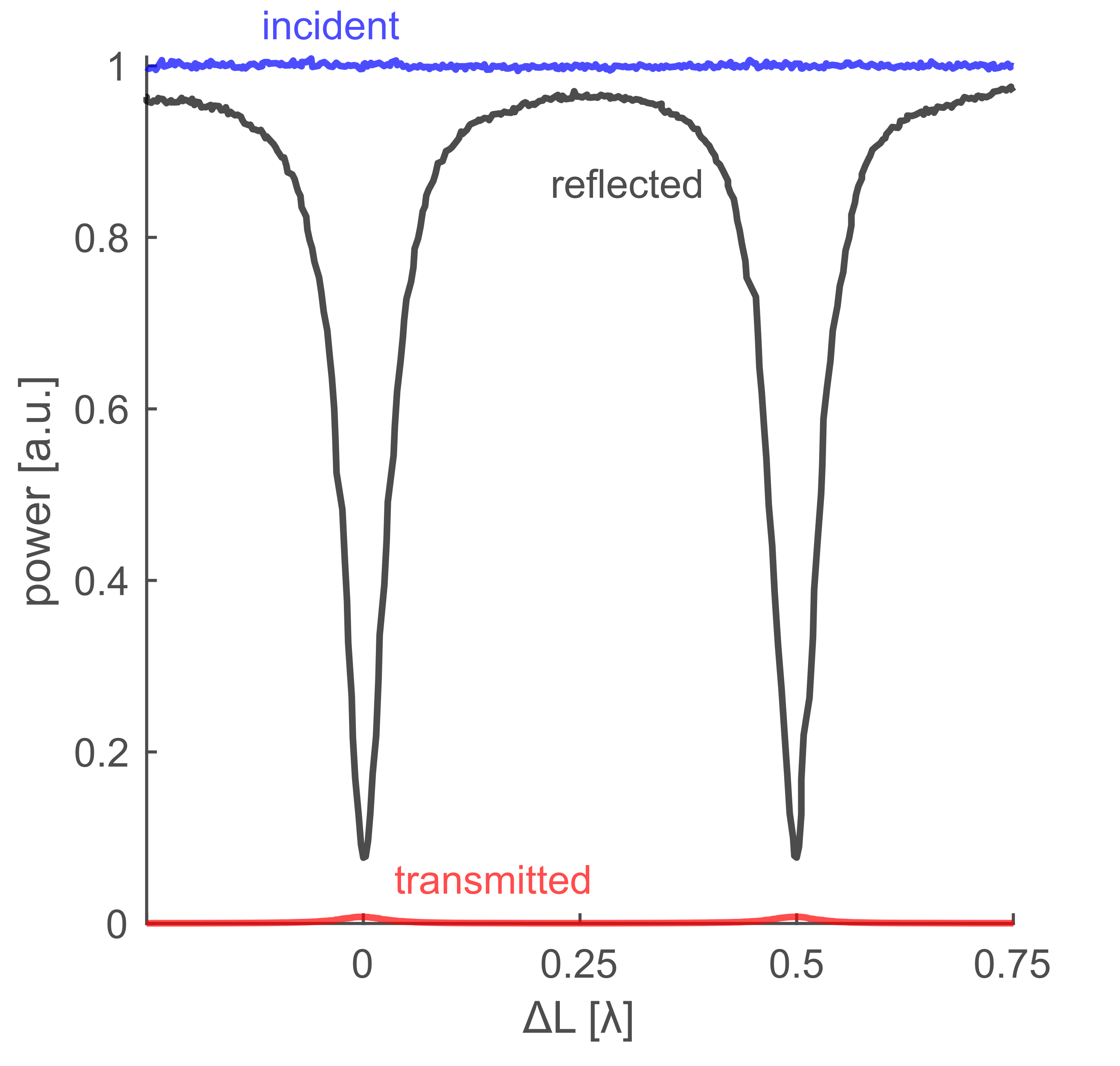}
\caption{\label{figS4:power_callibration} \textbf{Simultaneous measurements of transmitted, reflected, and total incident power.} The total incident (blue line), reflected (black line), and transmitted (red line) powers (normalized by the mean incident power) as measured while scanning the cavity length. The maximum power transmitted through the cavity was $<0.8\%$ of the total incident power, as theoretically expected.
}
\end{figure}

To validate that the non-reflected power is indeed absorbed and not resonantly transmitted through the cavity back mirror, we have measured the transmitted power through the rear cavity mirror when the cavity length was scanned. A comparison between the total incident, reflected, and transmitted powers, normalized by the mean incident power, is presented in Fig.~\ref{figS4:power_callibration} for an input field similar to the Yin-Yang pattern presented in Fig.~\ref{fig2:Setup_and_Ying_Yang}b of the main text.
The maximum power transmitted through the cavity back-mirror was $<0.8\%$ of the total incident power, as theoretically expected.

\end{document}